# Reversible photo-switching optical functionality in two-dimensional mixed-halide hybrid perovskites

Enamul Haque[1], Wenxin Mao[2], and Javier Cerrillo[1]

[1]Área de Física Aplicada, Universidad Politécnica de Cartagena, 30202 Cartagena, Spain

[2]Department of Chemical and Biological Engineering, Monash University, Clayton, Vic 3800, Australia

**Abstract**

Ion migration in halide perovskites is often associated with defects, irreversible processes, and structural instability, making them largely impractical for photo-switching applications. Here, we demonstrate a physical mechanism for reversible, defect-free light-induced halide-ion swapping in two-dimensional (2D) mixed-halide perovskites. We find that the halide-ion swapping process arises from strong light-lattice coupling rather than defects. By combining nudged elastic band (NEB) and photo-force calculations, we show that photo-induced forces perform non-equilibrium work that drives halide ions along the halide-exchange path without reaching the fully swapped configuration. Thus, the cumulative light-induced work can only partially overcome the ground-state activation energy barrier in the presence of light. Analysis of lattice dynamics identifies a few soft phonon modes, two of which are IR-active with oscillator strength ~-0.14 $e/(amu)^{1/2}$, which may be considered the microscopic origin of light-induced halide-ion swapping. This microscopic origin is further supported by band-edge–selective electron–phonon coupling, which amplifies interactions among excited carriers under illumination and with halide-ion motion without inducing a uniform dynamical instability. Using GW (G—Green's function and W—screened Coulomb interaction) calculations, we accurately reproduce the experimentally observed optical absorption spectra in the absence of light, enabling us to describe the light-induced excited state reliably. We demonstrate a clear redshift in the optical spectra in the presence of light, which is due to light-induced bandgap renormalization. Overall, these findings not only establish an intrinsic, defect-free mechanism for photoswitchable optical functionality in 2D mixed-halide perovskites but also demonstrate an intrinsic self-resetting feature in the absence of light.

## Introduction

Two-dimensional metal halide perovskites (2DHPs) are a class of crystalline materials with metal-halide octahedral layers separated by bulky organic cations to stabilize the structure and control interlayer spacing [1-4]. 2DHPs have gained significant interest in solar cells for their improved moisture stability and can be used as capping or photon-absorbing layers. [5-7]. The unique hybrid structure of 2DHPs enables strong quantum confinement and large exciton binding energies, making them suitable for LEDs, luminescent solar concentrators, and tunable lasers [2,3,6-8]. Similar to three-dimensional (3D) perovskites, the bandgap of 2DHPs can be tuned by controlling the halide composition [9] or by exposing them to light [10,11], further expanding their potential for advanced optoelectronic applications [7,12]. Exposing 2DHPs to light at an optimal wavelength can modulate their optical bandgaps [12]. Such light-controlled optical modulation, often called photoisomerization, is a crucial property for practical applications in photonic switching and optical sensing [8,13]. Consequently, photoisomerization triggered by light has gained significant interest due to its ability to induce substantial functional and structural changes [14]. However, in many halide perovskites, photo-induced halide ion motion is strongly coupled to irreversible ion migration, defect formation, and long-term ionic degradation [10,12,15,16], severely limiting its practical applications [12,15,17]. For practical applications, photo-induced ionic motion must be intrinsically reversible and efficient, so that it can enable repeatable optical modulation without causing electronic damage or drift.

Recently, researchers have successfully fabricated mixed-halide (Br/I) 2DHPs using a space-confining method and experimentally observed a unique halide isomerization of 2DHPs under photoexcitation, in which $Br^-$ and $I^-$ ions prefer to occupy bridging and terminal positions in the $PbX_6^{4-}$ octahedra, respectively [14]. Although the spatial separation of octahedral layers in 2DHPs limits halide species migration across inter-planes, halide exchange can still occur under photoexcitation or thermal excitation due to high concentration differences (i.e., a concentration gradient caused by photogenerated carriers) [14,15,18], even without causing macroscopic phase segregation or long-range ionic migration [14,15,18]. The study further demonstrated that the halide ions switching leads to reversible bandgap reduction and forms chemically distinguishable and activatable (through light) isomers. Such structurally encoded, site-selective, unique isomerization in 2DHPs opens up potential schemes for applications in quantum optoelectronic, sensing, and photonic devices [10,12,19,20]. While most photo-lattice effects are due to degradation, drift, and hysteresis, Mao et al. reported that such intrinsic, reversible photoisomerization is attributed to electronic excitation [14]. However, it remains unclear whether photoisomerization can also be attributed to intrinsic light-lattice coupling (including soft anharmonic-mode-induced coupling) or can intrinsically enhance photo-switchable functionality in 2DHPs.

Here, we demonstrate that light-lattice coupling can strongly contribute to the reversible photoisomerization in 2DHPs by using density functional theory (DFT) and GW (G—Green's function and W—screened Coulomb interaction) based first-principles calculations. We also show that photoisomerization intrinsically enables photoswitchable optical functionality, i.e., a reversible change in optical absorption upon light exposure. Based on nudged elastic band analysis and photo-force calculations, we show that light-lattice coupling generates direction-dependent forces that can transiently move halide ions along an intrinsic halide exchange path. Further analysis of the photo-induced cumulative work reveals that light can induce large-amplitude ionic motion, even without stabilizing the full exchange path, suggesting self-resetting functionality of the isomers upon light withdrawal. We further demonstrate the presence of a non-uniform e-ph coupling enhancement and the persistence of the ground-state barrier, both of which together prevent the energetic stabilization of the fully swapped structure. By correlating this light-lattice-coupling-induced halide-ion switching, which changes optical absorption, our results establish a clear understanding of a defect-free process for photo-switchable optical modulation in 2DHPs.

## Methods

We carried out the DFT calculations using the projector augmented wave (PAW) method as implemented in Vienna Ab initio Simulation Package (VASP) [21]. We used the PAW pseudopotentials to account for the effective couplings between valence and frozen-core electrons. To account for the exchange-correlation term, we utilized the generalized gradient approximation (GGA) of Perdew–Burke–Ernzerhof (PBE) functional [22,23]. We used a 400 eV cutoff energy and 12×12×1 -centered k-points for electronic-structure calculations. In these calculations, we used $10^{-8}$ eV for energy and 0.01 eV/Å for force convergences. Because the PBE functional severely underestimates the electronic band gap, we computed the electronic band structures using the $G_0W_0$ method [24-26]. In these calculations, we used 2×2×1 Γ-centered k-points and 315 eV cutoff energy. Then, we calculated the optical properties by solving the Bethe-Salpeter equations (BSE) [27-29]. We used VASPKIT [30] and AMSET [31] for post-processing some of the VASP data. We used VESTA for visualization of the crystal structures [32]. We calculated activation energy by using the nudged elastic band (NEB) method [33,34]. In these calculations, we set 4 × 4 × 1 the k-point and force convergence to 0.06 eV/Å. All NEB data were post-processed by using the VTST tool [34].

## Results and discussion

$BA_2PbBr_xI_{4-x}$ (BA = butylammonium, x = 4, 3, 2, 1, 0) crystallize in primitive orthorhombic crystal structures with four formula units per unit cell and Pbca space group [14,35], where BA acts as the spacer. Earlier studies have experimentally verified that larger iodine ions strongly prefer T-site (terminal site) occupancy, whereas smaller bromine or chlorine ions prefer the B-site (bridge site), as determined from single-crystal X-ray diffraction data [36,37]. Occupancy analysis also revealed that I has a strong preference for the terminal (T)-site, while Br is preferentially located at the bridge (B)-site [14]. Although other perovskites with mixed halide compositions have been reported, this specific arrangement of halides, with such selection of T-site and B-site, appears to be unique, as reported in Ref. [14].

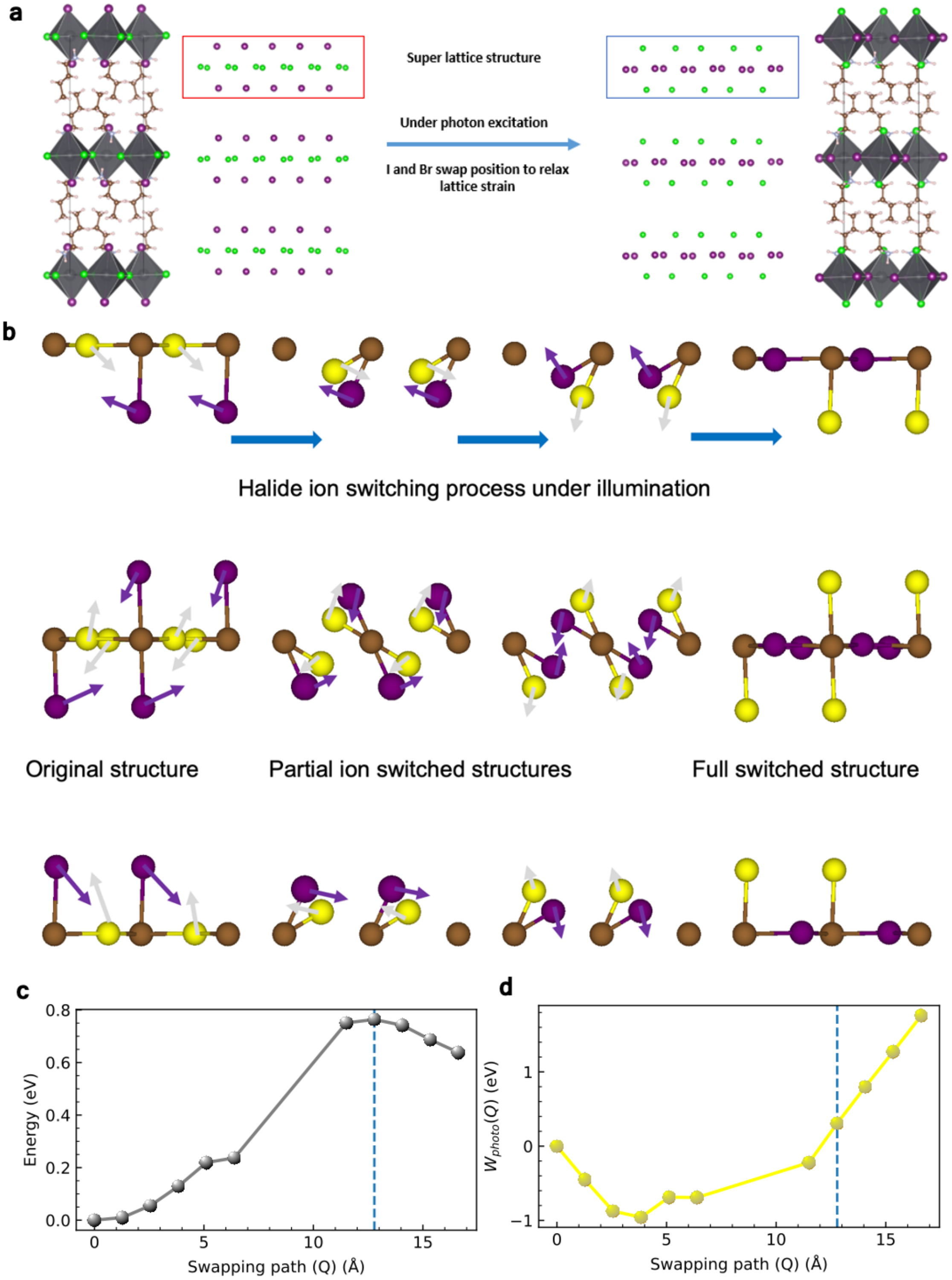

**Figure 1. Demonstration of energy profile and photo work for halide ions swapping.**

(**a**) Schematic diagram of light-induced halide ions switching, where a fully swapped structure has been shown for clarity. (**b**) Halide ions swapping path, i.e., exchange path for halide ions swapping in the presence of light. (**c**) NEB energy profile for halide ions swapping in $BA_2PbBr_2I_2$ in the absence of dark. The maximum height in the figure represents the activation barrier for full halide-ion swapping. (**d**) Photoinduced work in $BA_2PbBr_2I_2$ under light illumination. The dashed line indicates the transition state along the exchange path, where the activation barrier is located.

As experimentally demonstrated, the same halide ratio leads to the highest population of mixed halide octahedra [14]; we will mainly focus on the same halide-ratio system, i.e., on $BA_2PbBr_2I_2$ as a representative compound of 2DHPs, to illustrate the physical mechanism of halide-ion switching under photo excitation. Similar to the unique halide occupancy preference, in the simulation, half of the I anions are located at the T-site and the other half of the Br cations are located at the B-site (**Fig. 1a**). Upon photoexcitation, the interatomic forces is changed due to the modification of the electronic density and screening, which further induces excited-state forces. Such an excited-state force (or photo-force) transiently displaces halide ions away from their ground-state equilibrium positions, thereby relaxing interatomic stress (**Fig. 1b**). An earlier study experimentally verified that these halide ions swap positions, as shown in **Fig. 1b**.

To quantify these halide ion swaps, we construct a path of ion swaps with minimum energy between symmetry-equivalent configurations using the NEB method. In the NEB method, a reaction coordinate denoted by $r$ describes a continuous sequence of swapped structures $\mathbf{R}(r)$. The sequence connects the initial and final configurations by minimizing the energy and force perpendicular to the path. The energy profile $E_0(r)$ resulting from the minimization is used to estimate the ion migration barrier. This also identifies the transition state $r_{TS}$ that corresponds to the saddle point within the path. Importantly, this ion swap coordinate describes an intrinsic collective ionic motion, which is free from defects. The coordinate further serves as a natural basis for characterizing non-equilibrium forces produced by photoexcitation [33,34].

However, the ground state $E_0(r)$ only stands for the static energy that controls halide ions' motion in the absence of light. Local minimum energy along $E_0(r)$ represents a mechanically stable ground state structure, while the maximum energy along $E_0(r)$ indicates the energy that separates the transition state from equivalent structures. The height of this maximum energy $E_0(r)$ determines the energy barrier that ions must overcome for swapping their positions under dark conditions. As shown in Fig. 1b, halide ion swapping requires overcoming an energy barrier of approximately 0.78 eV to achieve full switching. Importantly, this $E_0(r)$ curve only represents the equilibrium energy path. It provides no information on external perturbations, *e.g.*, optical excitation. To overcome this activation barrier, it is pertinent to consider light as the primary driver of halide-ion movement, since photoexcitation can generate charge carriers, induce internal electric fields, alter defect charge states, and cause local photothermal heating. Ambient thermal energy at room temperature also facilitates this process. Specifically, kBT is approximately 0.026 eV at 300 K, which remains significantly lower than the activation barrier. Nonetheless, thermal lattice vibrations enable repeated hopping attempts, thereby allowing a small fraction of ions to surmount the barrier. When the combined energy contributions from light, thermal, and photothermal sources are inadequate, not all ions can surpass the 0.78 eV barrier, thereby inhibiting complete (100%) halide-ion switching.

To quantify the optical excitation effect on the halide ions swapping in the presence of light, we calculated the light-induced mechanical force that does the required work for ion swapping. In this case, the $BA_2PbBr_2I_2$ can no longer change based on $E_0(r)$, rather it experiences mechanical forces that drive the halide ions' motions. For a fixed ionic configuration $\mathbf{R}(Q)$, the photo-induced force can be defined as [38] $\Delta\mathbf{F}(r) = \mathbf{F}_{\text{photo}}(r) - \mathbf{F}_{\text{dark}}(r)$, where $\mathbf{F}_{\text{photo}}$ and $\mathbf{F}_{\text{dark}}$ represent the ionic forces computed in the presence of light (excited state) and under dark (ground state) conditions, respectively. Projecting this force onto the NEB tangent $\hat{T}(r) = \partial\mathbf{R}/\partial r$ produces the effective force that can drive the ions along the swapping path. By defining projectile force as $\Delta F_{\parallel}(r) = \sum_i \Delta\,\mathbf{F}_i(r)\cdot\hat{T}_i(r)$, the cumulative change by the photo-induced forces along the ions swapping path can be calculated by using the following equation [39-41]:

$$W_{\text{photo}}(r) = -\int_0^r \Delta\,F_{\parallel}(\mathrm{r})\mathrm{dr}.$$

This physical quantity measures the amount of mechanical work required to induce the motion of halide ions in the presence of light. A negative value of $W_{\text{photo}}(r)$ indicates total photo-driving force toward larger ionic displacement, while a positive $W_{\text{photo}}(r)$ indicates the barrier toward further motion along the stable ground state path. As shown in **Fig.1d**, the $W_{\text{photo}}(r)$ initially decreases $|W_{\text{photo}}(r_{TS})| > 0$with increasing $r$, indicating that photoexcitation carries out total mechanical work. This mechanical work pushes halide ions along the swapping path (**Fig. 1b**), thereby promoting ion motion. The minimum of $W_{\text{photo}}(r)$ at 3.0 Å stands for a structural configuration where the photo forces can balance. This configuration corresponds to a preferred photo-induced stable lattice distortion. Beyond this 3.0 Å distance along the path, $W_{\text{photo}}(r)$ starts to increase and then becomes positive above 7 Å. This demonstrates that further halide ion movement along the ion-swapping path is resisted in the presence of light, i.e., photo-induced forces cannot move the halide ions further along the swapping path of the ground state (**Fig. 1d**). More precisely, the effective energy required to create ionic movement can be defined as $E_{\text{eff}} = E_{NEB-barrier} - |W_{\text{photo}}(Q_{TS})|$,where $E_{NEB-barrier}$=0.78 eV, and $|W_{\text{photo}}(Q_{TS})|$= 0.31 eV. This equation indicates $0 < E_{\text{eff}} \ll E_{NEB-barrier}$, characterizing two scientific facts: First, the halide ions swapping process is a photo-assisted process, *i.e.*, additional thermal energy is required to achieve the fully swapped state. This clarifies why the previous experimental study only observed partial halide ion switching. This is also consistent with the previous experimental study, in which no halide-ion switching was observed at temperatures $< 200K$, supporting the idea that thermal energy is also critical for triggering switching. However, we anticipate that we may not observe any switching at high temperatures, where switching and reversing might be competing processes that partially cancel each other. Besides, the positive value of $W_{\text{photo}}(r)$ above 7 Å demonstrates that the photoexcited state cannot stabilize the fully ion-swapped structure (transition state) along the ion-swapped path. This behavior implies that photoexcitation can promote a transient halide-ion swap rather than irreversible diffusion. This also suggests that the halide ions return to their original positions when the light is removed, i.e., light illumination leads to reversible halide ion swapping that restores the same structural geometry as in the dark state.

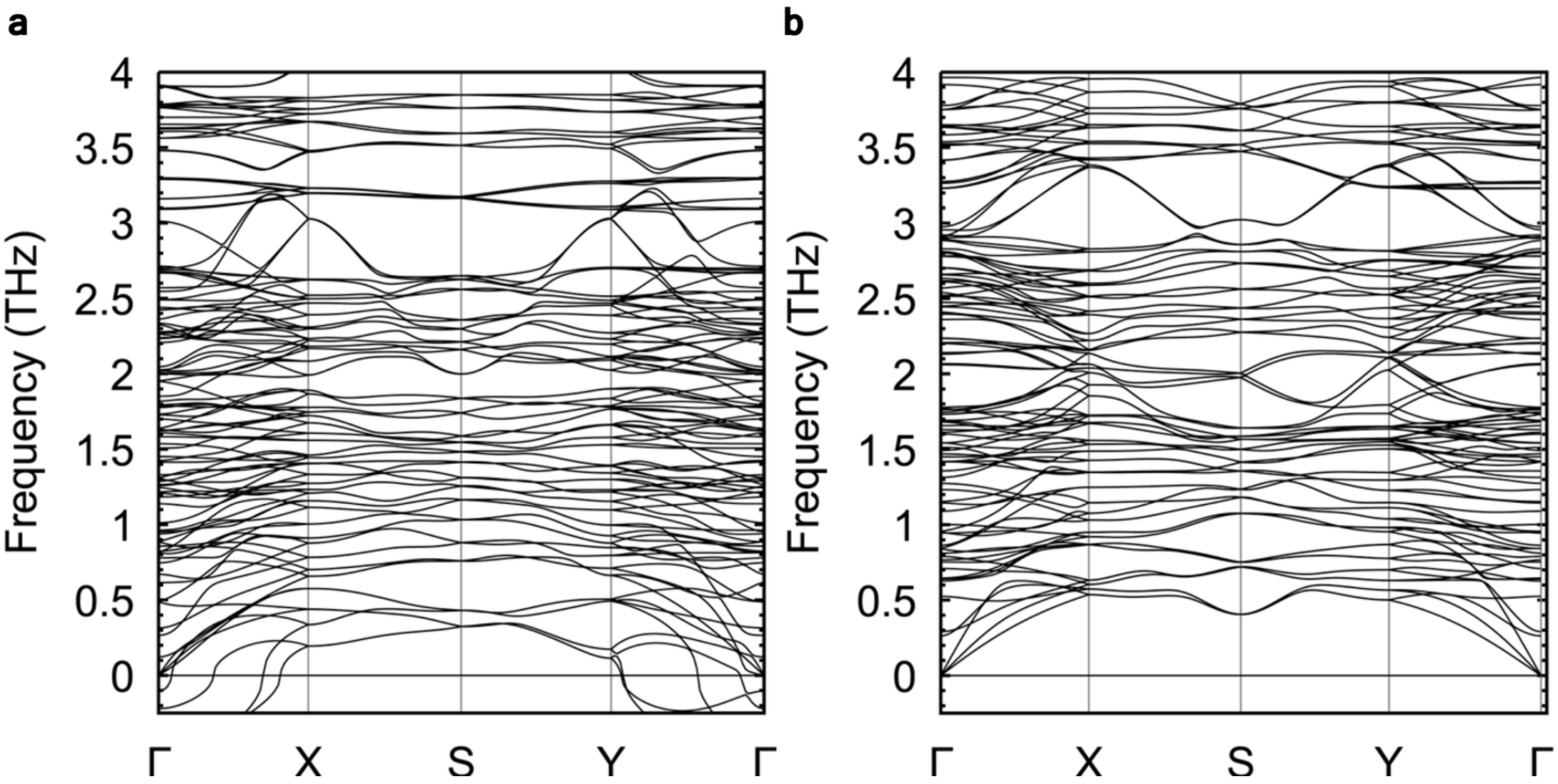

**Figure 2. Low-frequency phonon dispersion of $BA_2PbBr_2I_2$ under conditions of illumination and without illumination.**

(a) Phonon dispersion of $BA_2PbBr_2I_2$ in the absence of light, calculated excluding the SOC effect. (b) Phonon dispersion of $BA_2PbBr_2I_2$ in the presence of light calculated without considering the SOC effect. Only low-frequency phonon modes have been shown in the figure for clarity.

To examine how such large halide ions movement occurs in the presence of light, we calculated phonon dispersion of $BA_2PbBr_2I_2$ (Fig. 2). As shown in the figure, phonon dispersion of the ground-state configuration of $BA_2PbBr_2I_2$ possesses four soft phonon modes at the the zone center with negative frequencies —two of them are IR-active modes with a mode effective charge $\approx -0.14\ \mathrm{e}/\sqrt{\mathrm{amu}}$. Since these systems are centrosymmetric and have a small dielectric permittivity ($\approx 3$), they cannot be polar. Instead, these mixed halides feature phonon modes with anisotropic dipole moments. As a result, the calculated polarizability of $BA_2PbBr_2I_2$, including the Clausius-Mossotti correction, is sufficiently large ($\approx 218\ \text{Å}^3$) for light-matter coupling. Additionally, these imaginary modes near the zone center suggest a dynamical instability of the symmetric ground-state configuration with respect to a collective ionic displacement. These imaginary modes are related to directions in structural configuration space. Along this space, the potential energy curvature on the surface has a negative value [Ref. [14]]. Such a negative value implies that this structural configuration can still reside at a local maximum or saddle point, but it cannot reside at a true minimum along the direction of that phonon mode. In some halide perovskites, soft phonon modes are generally related to collective ion movements or octahedral ionic distortions. It reflects the intrinsic softness of these materials' lattices. Importantly, the presence of these soft modes in $BA_2PbBr_2I_2$ further reflects that they can induce a restoring force, or that the system feels internal stress (or strain), making it highly responsive to external perturbations, such as photons. However, the presence of these modes does not imply a spontaneous phase transformation at finite temperatures [42,43].

As we defined before, the halide ions' switching path corresponds to the ground-state NEB path. The lattice instability for these soft modes should evolve along the path. Intriguingly, despite having soft modes in the NEB ground state configuration, these soft modes disappear at the fully swapped (transition state) structure (**Fig. 2b**). All the phonon modes in the fully swapped configuration have positive frequency, reflecting its dynamical stability relative to halide ions displacements that are orthogonal to the ions' swapping path. This mechanical behavior is consistent with the NEB energy profile, in which we identified the first-order saddle point as the NEB saddle point, and the dynamical instability disappears along the halide-ion-swap path, i.e., it does not remain as a transverse dynamical instability. Therefore, the disappearance of these soft phonon modes in $BA_2PbBr_2I_2$ in the presence of light signifies a progressive hardening of the dynamical lattice as it approaches the fully halide ions swapped state [44].

This change in dynamical stability can microscopically explain the observed non-monotonic behavior of the photo work (**Fig. 1d**) along the ions-swap path described earlier. In particular, the presence of soft phonon modes in the ground-state configuration enables strong interactions between light and lattice, which is sufficiently strong to move halide ions along the swap path, producing a negative value of the (cumulative) photo work (**Fig. 1d**). As the halide ion movements are developed and the soft phonon modes have disappeared, the excited-state forces in the presence of light weaken along the ions swap path during the further ion movement. The excited-state forces eventually reverse sign along the swap path, increasing the photo work, and become positive near the fully swapped configuration. This coincidence between the disappearance of the dynamical instability and the change in the sign of photo work characterizes the fact that photoexcitation indeed stabilizes the fully halide ions-swapped structure itself, rather than the partially halide ions-swapped configuration, thereby relaxing the internal stress. This interplay between soft phonon modes and photo-excited-state forces also reflects the reversibility of halide-ion swapping in the presence of light observed in $BA_2PbBr_2I_2$. Explicitly, light illumination induces large halide ions movement, while the restoration forces of dynamical stiffness near the fully swapped structure create a barrier against irreversible halide ions migration, so that the halide ions can return to their original position upon removal of light illumination [44]. Similar to the ion swapping in $BA_2PbBr_2I_2$, phonon dispersions significantly evolve with mixing the halide ions or changing the mixing ratio of the halide ions, as shown in **Fig. 2** in the Supporting Information (SI).

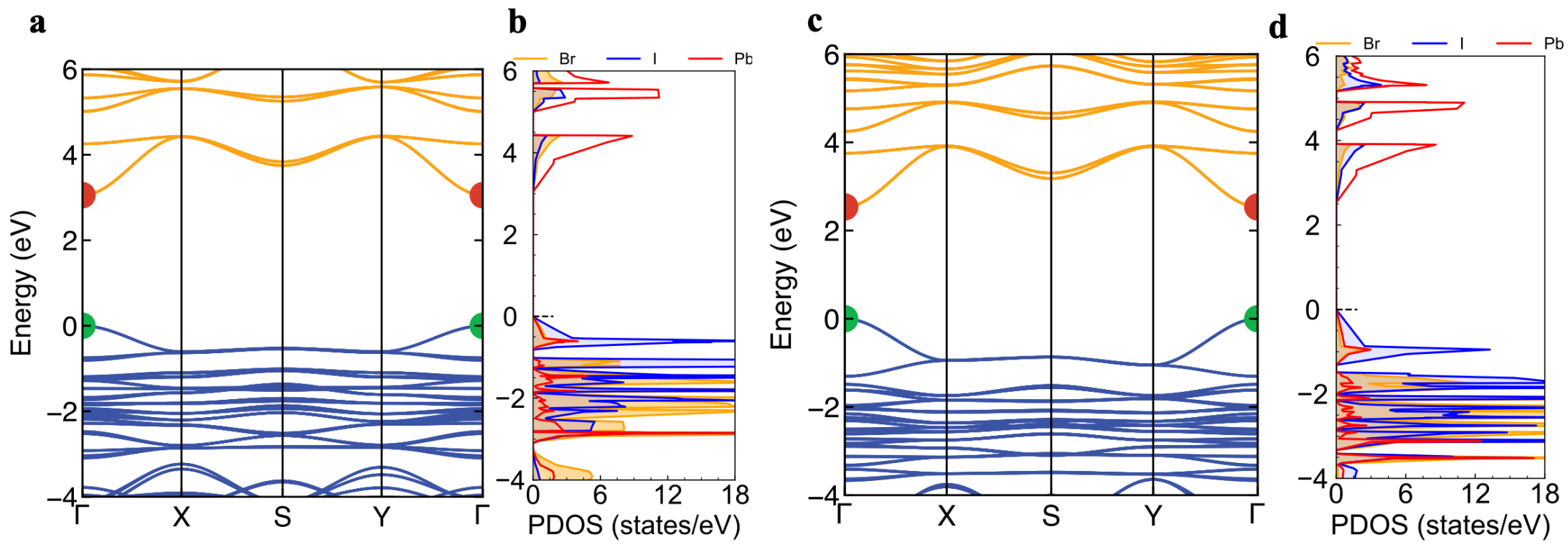

**Figure 3: Electronic structures (by using GW+SOC) of $BA_2PbBr_2I_2$. a**, Electronic band structure of $BA_2PbBr_2I_2$ over two-dimensional Brillouin zone (BZ) by using GW method. **b**, PDOS of $BA_2PbBr_2I_2$. **c**, Electronic band structure of $BA_2PbBr_2I_2$ in the presence of light (labelled as $BA_2PbI_2Br_2$). **d**, PDOS of $BA_2PbBr_2I_2$ in the presence of light (labelled as $BA_2PbI_2Br_2$).

Since these halide ion swaps strongly relate to energy and dynamical stability, we expect the corresponding electronic band structures to be largely influenced. Since the optical bangdap changes have already been explained theoretically and experimentally in our recent work [14,16]. we here focus on revealing the correlation between photo-assisted halide ion swapping and electronic structures. As suggested in the previous study [14], the electronic band structure and element-decomposed density of states indicate a pronounced change in the bandgap at the fully halide ions swapped state compared to the ground state (**Fig. 3a–c**). This narrowing of the gap is attributed to the substantial movement of halide ions inherent to the fully swapped structures. Such movement alters the Pb–halide bonding network. In particular, the element-projected density of states exhibits an enhanced hybridization between halide (Br or I)$-p$ states and Pb s orbital electrons near the band edges at the transition state, pushing up the valence-band maximum. Such hybridization also modifies the valence-band charge distributions, as suggested in a previous study [14]. Our calculated optical bandgaps and exciton binding energies for all systems (**Table 1**) using the GW+SOC method are in excellent agreement with experimental optical bandgaps [14]. However, as usual, the PBE functional severely underestimates the electronic bandgap. Besides, halide ion swapping also significantly reduces the hole effective mass (**Table 1**), suggesting improved optoelectronic performance.

**Table 1. Calculated electronic parameters of $BA_2PbBr_xI_{4-x}$ ($x$=0,1,2,3,4)**

Calculated fundamental bandgap ($E_g$), optical bandgap ($E_{opt}$), exciton binding energy ($E_{bind}$), electron ($m_e^*$) and hole ($m_e^*$) effective mass (in unit of bare electron mass, $m_0$), and deformation potential ($D$) (by using PBE functional). Usually, the fundamental bandgap ($E_g$) should be larger than the optical bandgap ($E_{opt}$) [45], as the excitation binding energies ($E_{bind}$) add to it, i. e. $E_{opt} = E_g - E_{bind}$. *BSE+SOC optical bandgap was not considered in the binding energy calculations as it severely underestimates the experimental values. **Bracket values of the binding energy were computed with respect to the experimental value [14]. ***$BA_2PbI_4$ structure at low temperature [14].

| Parameter | Method | System | | | | | | |
|---|---|---|---|---|---|---|---|---|
| | | $BA_2PbBr_4$ | $BA_2PbBr_3I$ | $BA_2PbBr_2I_2$ | $BA_2PbI_2Br_2$ | $BA_2PbBrI_3$ | $BA_2PbI_4$ | $BA_2PbI_4$*** |
| $E_g$ (eV) | PBE | 2.57 | 2.49 | 2.40 | 2.19 | 2.25 | 2.14 | 2.22 |
| | PBE+SOC | 1.87 | 1.78 | 1.71 | 1.45 | 1.55 | 1.42 | 1.60 |
| | GW | 4.02 | 3.92 | 3.80 | 3.28 | 3.48 | 3.34 | 3.39 |
| | GW+SOC | 3.24 | 3.15 | 3.05 | 2.53 | 2.72 | 2.50 | 2.73 |
| $E_{opt}$ (eV) | BSE | 2.88(2.95) | 2.85 (2.63) | 2.79 (2.53) | 2.34 (--) | 2.51(2.38) | 2.36 (2.27) | 2.51 (2.27) |
| | BSE+SOC* | 2.07 | 2.04 | 2.01 | 1.55 | 1.74 | 1.59 | 1.84 |
| $E_{bind}$ (eV) | GW** | 1.14(1.07) | 1.07(1.29) | 1.01(1.27) | 0.94(--) | 0.97(1.1) | 0.98(1.07) | 0.88(1.12) |
| | GW+SOC | 0.36(0.29) | 0.3(0.52) | 0.26(0.52) | 0.19 (--) | 0.21(0.34) | 0.14(0.23) | 0.22 (0.46) |
| $m_e^*$ $(m_0)$ | PBE | 0.323 | 0.309 | 0.303 | 0.301 | 0.299 | 0.270 | 0.159 |
| | PBE+SOC | 0.164 | 0.167 | 0.161 | 0.154 | 0.155 | 0.145 | 0.172 |
| | GW | 0.148 | 0.154 | 0.138 | 0.157 | 0.142 | 0.130 | 0.142 |
| | GW+SOC | 0.148 | 0.152 | 0.145 | 0.141 | 0.141 | 0.131 | 0.155 |
| $m_h^*$ $(m_0)$ | PBE | -0.314 | -0.358 | -0.408 | -0.234 | -0.300 | -0.270 | -0.509 |
| | PBE+SOC | -0.326 | -0.605 | -0.589 | -0.314 | -0.444 | -0.392 | -0.562 |
| | GW | -0.216 | -0.298 | -0.305 | -0.207 | -0.250 | -0.229 | -0.305 |
| | GW+SOC | -0.217 | -0.328 | -0.333 | -0.209 | -0.262 | -0.237 | -0.319 |
| D (eV) | PBE | 0.54 | 0.52 | 0.63 | 1.03 | 0.84 | 0.83 | -- |

Importantly, the bandgap renormalization in the presence of light directly describes the enhanced interactions between photoexcitation and halide-ion motion observed under illumination (as explained earlier using photoexcitation-induced work). As the halide ions swap along the exchange path, the bandgap renormalization should also alter the electronic polarizability and, hence, amplify the photoexcitation response in the electronic density. Such a strong response enhances the forces driving ion swapping acting on the halide ions in the presence of light, leading to large photo-driven halide-ion motion. Despite bandgap renormalization, however, the fully swapped state cannot become stable in the presence of light, as suggested by the photo-driven cumulative work.

The relations among bandgap renormalization, the disappearance of phonon soft modes, and the reversal of photo-driving under light reveal the fundamentally excited-state nature of the photoexcitation response in halide ions. At the same time, the dynamic stabilization and the persistence of positive induced work beyond the transition point along the ions swap path create a barrier against the energetic stabilization of the fully swapped structure. As a result, photoexcitation provides a transient pathway to access electronically soft, highly polarizable structural configurations in the presence of light, even without inducing irreversible ionic diffusion. Thus, such electronic softening in the fully swapped structure provides a microscopic connection between halide-ion motion, optical absorption, and reversible photo-induced ion swapping in wide-bandgap 2D mixed-halide perovskites [44]. We also observed similar bandgap renormalization upon introducing I into the system and increasing its content (**Fig. S3** in the SI), consistent with previous studies. This unique feature suggests that the bandgap renormalization may significantly modify the electron-phonon interactions without overcoming the energy barrier along the ion-swapping path [44]. The fully swapped structure may correspond to a configuration of maximal electron-phonon interactions, in which bandgap renormalization due to strong orbital hybridization leads to efficient photo-lattice coupling. This is well evidenced from the calculated deformation potential (**Table 1**), where a significant increase of its value indicates a much stronger electron-phonon scattering in the fully swapped configuration, suggesting that the lattice is highly susceptible to collective halide ions movement [33,43,44].

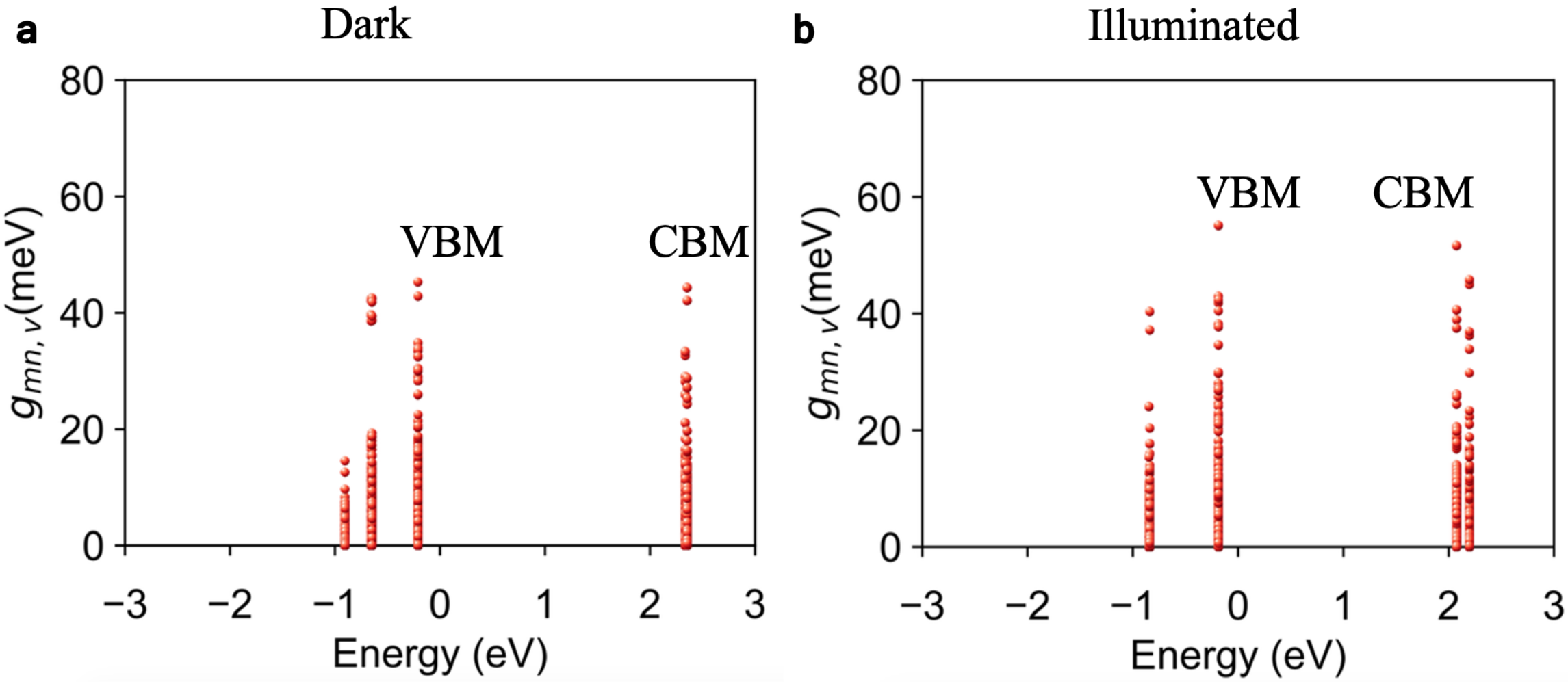

**Figure 4. Electron–phonon interaction matrix as a function of electronic energy of $BA_2PbBr_2I_2$.**

Electron–phonon (e–ph) matrix elements as a function of electronic energy for (a) the ground-state structure of $BA_2PbBr_2I_2$ in the absence of light and (b) the fully halide-swapped structure of $BA_2PbI_2Br_2$ under illumination. In both subfigures, we label the valence-band maximum (VBM) and the conduction-band minimum (CBM). Each sphere represents the magnitude of the e–ph matrix element for electronic states within a selected window around the band edges and for the corresponding phononic states.

To quantify this e-ph interactions further, we calculated the electronic band-resolved electron-phonon (e-ph) coupling matrix via density functional perturbation theory using the equation,[46,47]

$$g_{mn,v}(\boldsymbol{k},\boldsymbol{q}) = \sqrt{\left(\frac{\hbar}{2M\omega_{\boldsymbol{q}v}}\right)}\langle\psi_{m\boldsymbol{k}+\boldsymbol{q}}|\partial_{\boldsymbol{q}v}V_{SCF}|\psi_{n\boldsymbol{k}}\rangle. \quad (1)$$

Here $\psi_{nk}$ ($\psi_{mk+q}$) represents the $n^{th}$ ($m^{th}$) Kohn-Sham wavefunction at electronic wavevector $\boldsymbol{k}$ ($\boldsymbol{k}+\boldsymbol{q}$) with electronic eigenvalue $\epsilon_{nk}$ ($\epsilon_{mk+q}$), $\omega_{qv}$ stands for the $v^{th}$ phonon energy at phonon wavevector $\boldsymbol{q}$, and $\partial_{qv}V_{SCF}$ is the first-order derivative of Kohn-Sham self-consistent potential[46,47]. We present the calculated e–ph matrix elements as a function of electronic energy for electronic states near the band edges (**Fig. 4**). The results reveal pronounced non-uniform variation in e-ph coupling with halide-ion motion. In particular, the e–ph coupling at the valence-band maximum (VBM) and conduction-band minimum (CBM) is substantially enhanced in the fully halide-swapped structure, whereas electronic states further below the VBM and above the CBM show reduced e-ph coupling compared to the ground-state structure in the absence of light (**Figure 4a)**. This demonstrates that halide-ion motion causes a non-uniform increase in e-ph interaction. However, halide-ion switching selectively strengthens e-ph coupling at the electronic band edges (**Fig. 4b)**.

The electronic band-resolved e–ph coupling further explains the origin of the non-monotonic mechanical photo-work along the exchange pathway. As the halide ions begin to move within the ground-state structure under illumination, the system enters a regime where the e-ph coupling to lattice motion strengthens at band edges, resulting in strong excited-state forces. These strong photo-forces drive ionic motion and produce negative mechanical photo-work. Importantly, this regime corresponds to the initial regime of efficient photo-driven halide ion motion. When the halide ions move further, the system approaches the transition-state structure. However, the photo-induced forces arising from e-ph coupling can no longer displace the halide ion further, thereby reversing the photo-induced work. In the context of optical excitation, such enhancement is highly significant because optical absorption mainly involves transitions between electronic states near the CBM and VBM. The enhanced e–ph coupling at the band edges in the fully swapped configuration suggested that photo-excited carriers feel stronger coupling to lattice motion when light drives the system away from equilibrium. On the other hand, the e-ph coupling weakens for electronic states away from the VBM and CBM. As the e–ph coupling is significantly enhanced at the band edges, rather than across all states, the photo-ionic response may be intrinsically connected to optically active states. This band-selective behavior of e-ph coupling provides a direct microscopic description of the observed halide-ion-motion-induced bandgap reduction and potential redshift in the absorption spectra.

To visualize a potential shift in the optical absorption spectrum, we calculated it using the GW method and electron–hole interactions within the Bethe–Salpeter equation (BSE). As shown in **Fig. 4a**, the optical absorption spectrum of $BA_2PbBr_2I_2$ shows excellent agreement with experimental absorption data in the absence of light [14] (i.e., equilibrium structure). Both the absorption onset and the main spectral features are accurately reproduced relative to the experimental data. This confirms well-captured behavior of the quasiparticle electronic band structures and excitonic effects. Such excellent agreement between computed and experimental data validates the use of the combined GW+BSE method to reliably describe photoexcited optoelectronic states relevant to our present study. It also establishes the method as a robust baseline for interpreting light-induced changes in the optoelectronic properties using intrinsic light-matter coupling rather than extrinsic defects, disorder, or diffusion.

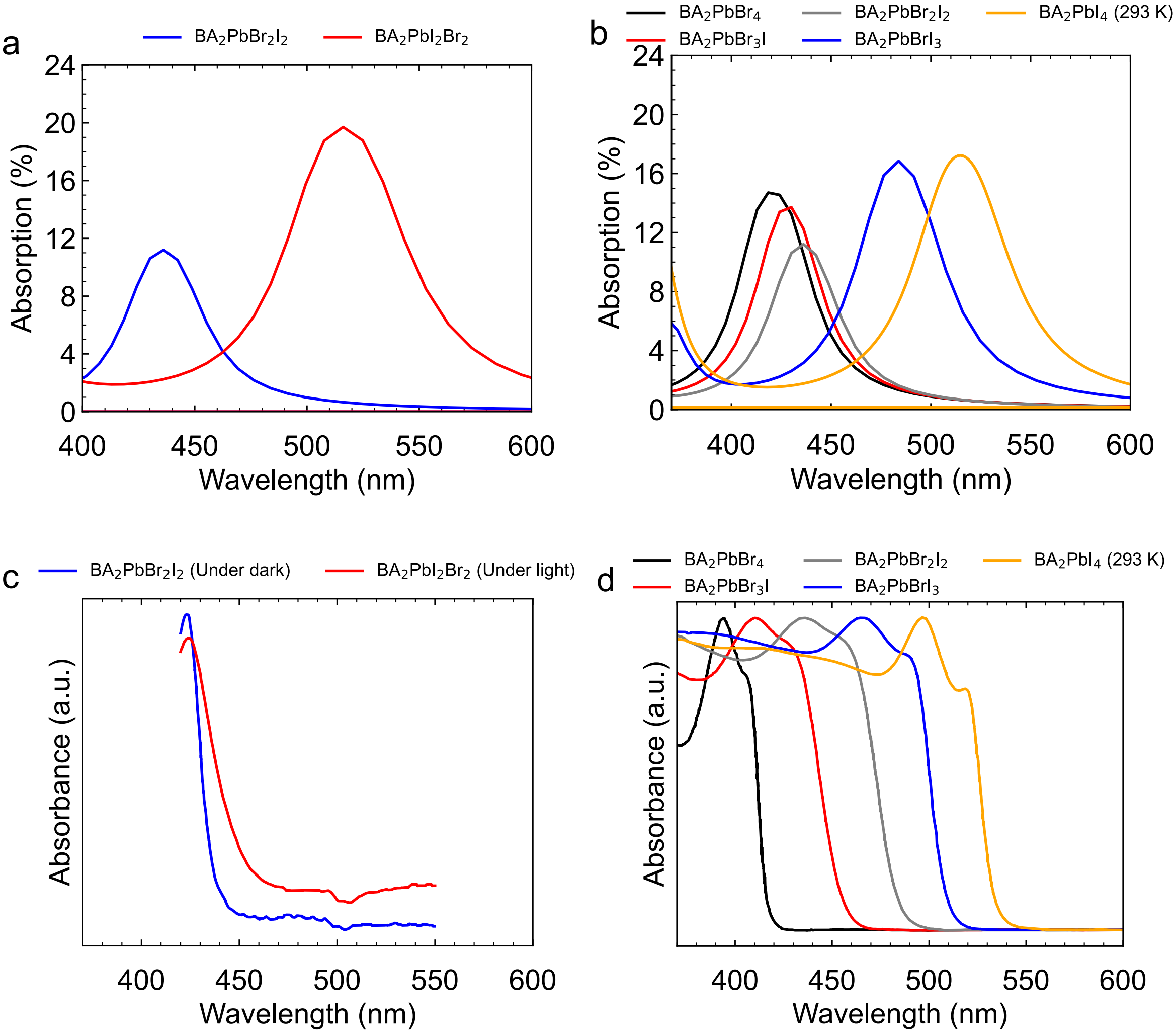

**Figure 5. Calculated optical absorption spectra of $BA_2PbBr_xI_{4-x}$ and one fully swapped configuration.**

(**a**) Optical absorption spectra of $BA_2PbBr_2I_2$ in the absence of light and in the presence of light (labelled as $BA_2PbI_2Br_2$ for a fully halide ion swapped configuration). (**b**) Optical absorption spectra of $BA_2PbBr_4$, $BA_2PbBr_3I$, $BA_2PbBrI_3$, and $BA_2PbI_4$ (This structure corresponds to the structure obtained at 293 K in the experiment [14]). (**c**) Experimentally measured absorbance spectra of $BA_2PbBr_2I_2$ in the absence of light and in the presence of light. These data were taken from Ref. [14]. (**d**) Our experimentally measured optical absorbance spectra of $BA_2PbBr_4$, $BA_2PbBr_3I$, $BA_2PbBrI_3$, and $BA_2PbI_4$. We used same experimental conditions and same measurement described in Ref. (14).

Upon halide-ion switching along the exchange path, the calculated absorption edge shows a clear redshift (**Fig. 5b**), consistent with the bandgap renormalization observed in the fully swapped and partially swapped structures. This redshift demonstrates halide-ion-switching-induced electronic softening and enhanced polarizability, thereby lowering the energy cost of photoexcitation. Importantly, the same halide-ion-swapped structures that show redshifts in the absorption spectra also correspond to regions where the photoexcitation-induced driving forces are strong enough and electron-phonon interactions are sufficiently enhanced (**Fig. 4**). This feature indicates a direct connection among absorption spectra, photoexcited-state forces, and halide-ion motion. The photon strongly couples to lattice vibrations of the original state, where the lattice is electronically softened (negative or reduced frequencies shown in **Fig. 2a** compared to **Fig. 2b**). Consequently, photoexcitation amplifies the photo-induced response, keeping the fully halide-ion-swapped states energetically unstable. Therefore, the redshift in the absorption spectra serves as a spectroscopic signature of the fully halide-ion-swapped lattice, underpinning the reversible photo-induced movement of halide ions. Compared with halide-ion swapping, mixing Br and I in the same system and increasing I content also yield a clear redshift in the optical spectra of all studied systems (**Fig. 5b**). However, halide-ion swapping can significantly improve the absorption of $BA_2PbBr_2I_2$ compared with halide mixing (e.g., $BA_2PbBr_3I$), suggesting that light-matter coupling can be a unique approach to enhance the optoelectronic performance of this family of materials studied here. These results indicate that the halide-ion-swapped structure corresponds to an electronically softened, optically more active lattice configuration, in which photoexcited carriers couple strongly to halide-ion switching. At the same time, the presence of a non-uniform e-ph coupling enhancement and the persistence of the ground-state barrier prevent the energetic stabilization of the fully swapped configuration, ensuring reversibility upon removal of illumination. Therefore, such non-uniform and band-selective e-ph coupling offers a unique microscopic approach to connect reversible halide-ion switching, photoexcited forces, bandgap renormalization, and optical absorption in 2D mixed-halide perovskites. The band-edge-selective nature of electron–phonon coupling therefore provides a unifying microscopic framework connecting optical absorption, excited-state forces, and reversible photo-ionic displacement in wide-bandgap double perovskites.

**Conclusions**

In summary, we computationally demonstrate a reversible and intrinsic photoexcitation-induced halide ions (partial) switching in 2D mixed halide perovskite. We also demonstrate that halide ion swapping is a photo-assisted process that arises primarily from light-soft phonon coupling rather than from defects or impurities. By combining nudged elastic band (NEB) and excited-state force calculations, we demonstrate that photoexcitation performs non-equilibrium mechanical work that drives halide ion movement toward a swap structure without stabilizing the fully halide-ion-swapped configuration. The calculated cumulative photo-induced work reveals partial assistance in overcoming the ground-state activation energy barrier, enabling photo-assisted but reversible halide ion swapping. Analyses of lattice dynamics identify four soft phonon modes—three of them are acoustic modes, among which two modes are anisotropic IR active modes, suggesting strong dipole-dipole interactions, which can be considered as the main microscopic origin of the light-matter coupling, while halide ions swapping induces a clear bandgap renormalization. Using the GW method and the Bethe–Salpeter equation (BSE), including electron–hole interactions and spin–orbit coupling (SOC), we accurately reproduce the experimentally observed optical absorption spectra in the absence of light. This establishes a reliable description of photoexcited transition states. The absorption spectrum of $BA_2PbBr_2I_2$ shows a clear redshift and a significant improvement in absorption under illumination. Our findings describe an intrinsic, diffusion-free mechanism for photoswitchable optical functionality in 2D mixed-halide perovskites, including an intrinsically self-resetting feature under ambient conditions. Overall, the above results clearly establish a defect-free, diffusion-free route to photo-switchable optical functionality in 2D mixed-halide perovskites, opening opportunities for highly stable, self-switching photonic and sensing devices.

**Acknowledgment**

E. H. and J. C. sincerely acknowledge financial support from the grant CNS2023-144994, funded by MICIU/AEI/10.13039/501100011033 and by “ERDF/EU,” and from the European Union project C-QuENS (Grant No. 101135359).

**Conflict of Interest**

There is no conflict of interest to declare.

**Data Availability**

All data are included in the manuscript, and the raw data are available upon request from the corresponding author.

# Reversible photo-switching optical functionality in two-dimensional mixed-halide hybrid perovskites

Enamul Haque[1], Wenxin Mao[2], and Javier Cerrillo[1]

[1]Área de Física Aplicada, Universidad Politécnica de Cartagena, 30202 Cartagena, Spain

[2]Department of Chemical and Biological Engineering, Monash University, Clayton, Vic 3800, Australia

## Supplementary Table 1

**Calculated lattice parameters and formation energy of $BA_2PbBr_xI_{4-x}$ ($x$=0,1,2,3,4) systems.**

| System | a (Å) | b (Å) | c (Å) | Formation energy (eV/atom) |
|---|---|---|---|---|
| $BA_2PbBr_4$ | 8.2430 | 8.1610 | 27.560 | -0.3294 |
| $BA_2PbBr_3I$ | 8.2660 | 8.1550 | 27.599 | -0.3161 |
| $BA_2PbBr_2I_2$ | 8.3970 | 8.2500 | 27.784 | -0.3027 |
| $BA_2PbI_2Br_2$ | 8.5766 | 8.3616 | 27.547 | -0.2993 |
| $BA_2PbBrI_3$ | 8.4880 | 8.3600 | 27.715 | -0.2898 |
| $BA_2PbI_4$ | 8.4340 | 8.9880 | 26.275 | -0.2749 |
| $BA_2PbI_4$* | 8.7573 | 8.5848 | 27.8828 | -0.2773 |

* $BA_2PbI_4$ structure at 200 K temperature. The structure is extracted from (ref. 14)

Note**:
The formation energy can be calculated by using the following equation:

$$H_f\ (eV) = E_{tot} - (4E_{Pb} + 8E_N + 96E_H + 32E_C + 16E_{Br/I}) \ldots\ldots (S1)$$

where $E_{tot}$ is the total ground state energy of $BA_2PbBr_xI_{4-x}$; $E_{Pb}$, $E_N$, $E_H$, $E_C$, and $E_{Br/I}$ are the ground state energy (per atom) of Pb, N, H, C, and Br/I in their corresponding stable structures. The calculated formation energy indicates that the swapped structure ($BA_2PbI_2Br_2$ labeled in red) is less energetically stable than the original structure, suggesting that it is in an excited state corresponding to the structure under illumination Note: the highlight represent the lattice constant and formation for the excited structure.

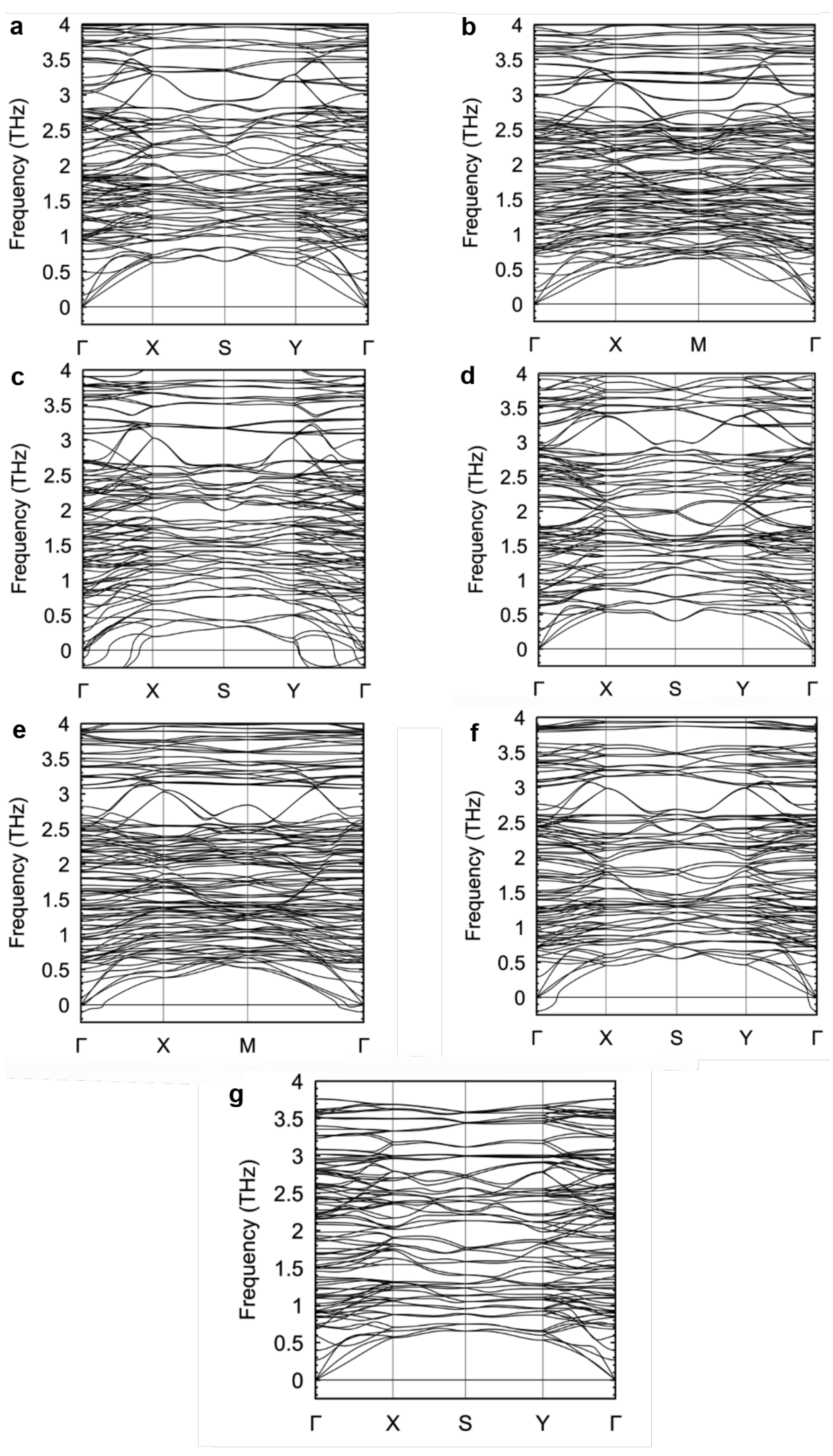

**Fig. S1. Phononic band structures (by PBE) of $BA_2PbBr_xI_{4-x}$ (*x*=0,1,2,3,4).**

**(a)** Phononic band structure of $BA_2PbBr_4$ over two-dimensional Brillouin zone (BZ) by using PBE functional. (**b**) Phononic band structure of $BA_2PbBr_3I$. (**c**,) Phononic band structure of $BA_2PbBrI_3$. (**d**) Phononic band structure of $BA_2PbBr_2I_2$. (**e**) Phononic band structure of $BA_2PbI_2Br_2$. (**f**) Phononic band structure of $BA_2PbI_4$. (**g**) Phononic band structure of $BA_2PbI_4$* (structure at 293 K).

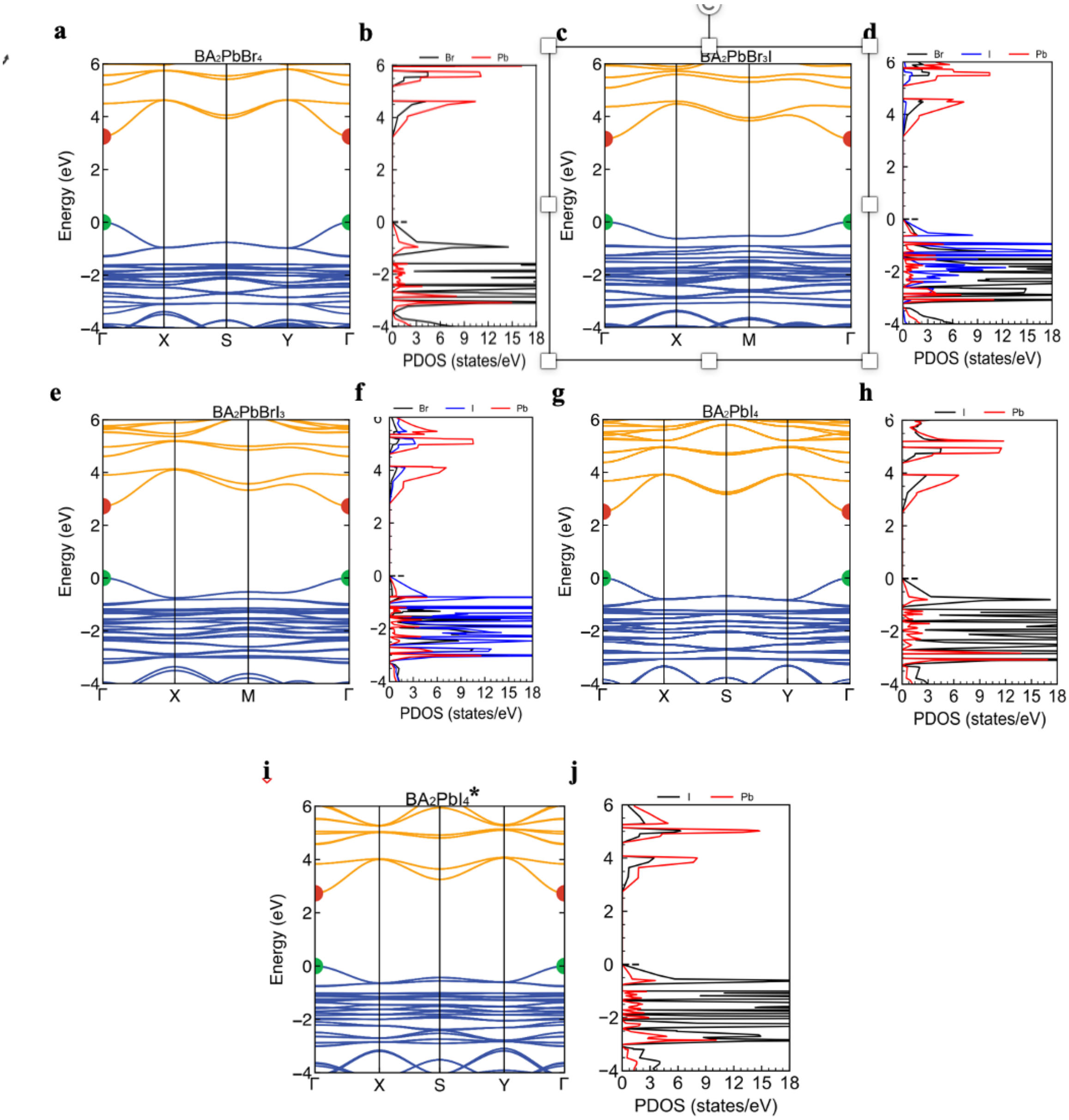

**Fig. S2. Electronic structures (GW+SOC) of $BA_2Pb(Br_xI_{1-x})_4$ ($x$=0,1,2,3,4).**

(**a**) Electronic band structure of $BA_2PbBr_4$ over two-dimensional Brillouin zone (BZ) by using GW method. (**b**) PDOS of $BA_2PbBr_4$. (**c**) Electronic band structure of $BA_2PbBr_3I$. (**d**) PDOS of $BA_2PbBr_3I$. (**e**) Electronic band structure of $BA_2PbBrI_3$. (**f**) PDOS of $BA_2PbBrI_3$. (**g**) Electronic band structure of $BA_2PbI_4$. (**h**) PDOS of $BA_2PbI_4$. (**i**) Electronic band structure of $BA_2PbI_4$* (structure at low temperature). (**j**) PDOS of $BA_2PbI_4$*